\def\year{2020}\relax
\documentclass[letterpaper]{article} %
\usepackage{aaai20}  %
\usepackage{times}  %
\usepackage{helvet} %
\usepackage{courier}  %
\usepackage[hyphens]{url}  %
\usepackage{graphicx} %
\usepackage{graphicx}  %
\nocopyright
\usepackage{amsfonts}
\usepackage{amsmath}

\usepackage[dvipsnames]{xcolor}

\newcommand{\citet}[1]{\citeauthor{#1} \shortcite{#1}}

\usepackage{pifont}
\usepackage{graphicx} %
\graphicspath{{plots/}} %

\usepackage{tablefootnote}

\usepackage{subcaption}

\usepackage{bm}
\usepackage[mathscr]{euscript}

\usepackage{color}
\usepackage{xspace}
\usepackage[]{todonotes}   %

\usepackage{algorithm,algorithmicx}
\floatname{algorithm}{Listing}
\usepackage[noend]{algpseudocode}
\MakeRobust{\Call}   %
\newcommand{\rightcomment}[1]{\(\triangleright\) {\footnotesize\textit{#1}}}
\algrenewcommand{\algorithmiccomment}[1]{\hfill \rightcomment{\normalsize #1}}  %
\algnewcommand{\LineComment}[1]{\State \rightcomment{#1}}
\algnewcommand{\LinesComment}[1]{\State \rightcomment{\parbox[t]{\linewidth-\leftmargin-\widthof{\(\triangleright\) }}{#1}}\smallskip}

\algrenewcommand\algorithmicindent{1.0em}%

\usepackage{tikz}
\usetikzlibrary{positioning,calc}
\usepackage{tikz-3dplot}
\usepackage{pgfplots}

\usepackage{varwidth}

\usepackage{cleveref}
\floatname{algorithm}{Listing}
\crefname{algorithm}{listing}{listings}
\Crefname{algorithm}{Listing}{Listings}
\crefname{section}{section \textsection\kern-.3ex}{sections}
\Crefname{section}{Section \textsection}{Sections}

\crefname{figure}{figure}{figures}
\Crefname{figure}{Figure}{Figures}
\crefname{equation}{equation}{equations}
\Crefname{equation}{Equation}{Equations}

\newcommand\blfootnote[1]{%
  \begingroup
  \renewcommand\thefootnote{}\footnote{#1}%
  \addtocounter{footnote}{-1}%
  \endgroup
}

\newcommand{\defn}[1]{\textbf{#1}\xspace}

\def\dim{\ensuremath{n}}

\newcommand{\norm}[1]{\ensuremath{\lVert #1 \rVert}}
\def\vecspace{\ensuremath{\mathbb{R}^\dim_{\norm{\cdot}=1}}}
\def\cosine{\text{cosine}}
\def\vertices{\ensuremath{\mathcal{V}}}
\def\lneighbor{\ensuremath{\Gamma}}  %
\def\constraints{\ensuremath{\mathcal{C}}}

\def\acos{\ensuremath{\cos^{\scalebox{.75}[1.0]{-}1}}}

\renewcommand{\v}[1]{v_{#1}}  %
\newcommand{\vi}{\v{i}}

\newcommand{\cc}{$\mathbf{C}_2$\xspace}

\newcommand{\shrinkDoc}[1]{}

\title{Exact and/or Fast Nearest Neighbors}

\author{Matthew Francis-Landau, Benjamin Van Durme \\
  Johns Hopkins, Department of Computer Science \\
  \{mfl,vandurme\}@cs.jhu.edu}

\begin{document}

\maketitle

\begin{abstract}
  Prior methods for retrieval of nearest neighbors in high dimensions are fast
  and approximate--providing probabilistic guarantees of returning the correct
  answer--or slow and exact performing an exhaustive search.  We present
  \emph{Certified Cosine} (\cc), a novel approach to nearest-neighbors which
  takes advantage of structure present in the cosine similarity distance metric
  to offer \emph{certificates}.  When a certificate is constructed, it
  guarantees that the nearest neighbor set is correct, possibly avoiding an
  exhaustive search.  \cc's certificates work with high dimensional data and
  outperforms previous exact nearest neighbor methods on these datasets.
\end{abstract}

\section{Introduction}

Abstractly, the nearest neighbor problem is defined as given a query
$q \in \mathbb{R}^n$, find the \emph{nearest} vector $\vi \in \mathcal{V}$ from
a discrete set of points according to a distance function $d(x, y)$
($\text{argmin}_{\vi} d(q, \vi)$).  Nearest neighbors occurs frequently as a
subproblem in document retrieval~\cite{embedding_document_retrieval}, image
search~\cite{faiss} and language
modeling/generation~\cite{Bengio:2003:NPL:944919.944966,sugawara-etal-2016-approximately}. Because
$\mathcal{V}$ is often very large, the time spent searching $\mathcal{V}$
dominates the time to evaluate and train a machine learning model.  Hence, this
has motivated the development of \emph{fast} nearest neighbor methods.%
\blfootnote{Source code available at: \\
  {\tt github.com/matthewfl/certified-cosine}}

\subsection{Prior Nearest Neighbor Methods}

Prior fast nearest neighbor (NN) methods fall into two main categories:
\emph{exact} and \emph{approximate}.  Exact methods, such as
KD-trees~\cite{kdtrees}, VP-tree~\cite{Yianilos:1993:DSA:313559.313789} and
cover-trees~\cite{cover_tree}, only work well in low dimensional settings (such
as graphics with 3-dimensions).  These methods work by first building an index in the form
of a tree data structure which at every level will split the data according to
some separating hyperplane.  The separating hyperplane may be chosen according
to the standard basis such as in KD-trees or a radial basis as in VP-trees.
When searching for the nearest neighbor of $q$, these methods first locate an initial
guess $\hat{v}$ by greedily searching their tree index.
Using $\hat{v}$, these methods will search branches which \emph{might}
contain a better neighbor $\norm{v_i - q} \le \norm{\hat{v} - q}$ and prune any
branch of the tree that provably does not contain anything better than
$\hat{v}$~\cite{kdtrees,cover_tree,Yianilos:1993:DSA:313559.313789,Ciaccia:1997:MEA:645923.671005,chen2018_convexhulltree}.

The difficulty with these exact methods in high dimensional settings---such as
used with machine learning methods---is that clustering of vectors into
different branches is incapable of eliminating regions of the search space.
Essentially, the distance between the \emph{nearest} neighbor and the
\emph{furthest} neighbor vanishes making it impossible to prune branches of the
search tree\footnote{Colloquially this problem of vanishing distances between
  different neighbors is known as the \emph{curse of
    dimensionality}.}~\cite{when_nn_is_meaningful}.

Being unable to prune branches of a search tree has motivated the development of
many \emph{approximate} nearest neighbor (ANN) methods for working with high
dimensional data.  Instead, these methods search/prune the space
\emph{probabilistically}.  Generally, these methods have an $\epsilon$ parameter
to trade-off the recall against the runtime and storage complexity
(E.g. $P(\hat{v} = v^*) \ge 1 - \epsilon$ with a runtime of
$O(\epsilon^{-O(1)})$).  The exact details for how search is performed and how
the parameter $\epsilon$ integrates varies significantly between different ANN
methods.  One common approach has been to use random projections to a
lower-dimensional space, such as used by Locality Sensitive Hashing (LSH) and
its
derivatives~\cite{Indyk:1998:ANN:276698.276876,Charikar:2002,faiss,Muja09fastapproximate,DBLP:conf/icml/LiM17,rpforest}.
With methods like LSH, vectors $\mathcal{V}$ are partitioned into ``hashed
buckets.''  When searching, the probability that a bucket contains the nearest
neighbor can be computed by comparing hashing difference between the bucket and
the query $q$.  This, in turn, is used to bound the probability that the nearest
neighbor is not found.  However, if an application requires that the
\emph{exact} nearest neighbors ($\epsilon = 0$), then bounding the probability
of not finding $v^*$ does not work.  In the exact case, these methods require
searching \emph{all} buckets and achieve no speedup.

Another class of approximate methods is based on the \defn{$K$-nearest neighbor
  graph}
(KNNG)~\cite{Arya:1993:ANN:313559.313768,Sebastian:2002:MSR:839291.842892}.  A
KNNG represents all vectors $\vi \in \vertices$ as vertices in the graph.  Edges
of the KNNG correspond with the $k$-nearest neighbors for every vector which is
computed once during preprocessing.  During search, regions that are \emph{near}
to the query $q$ are prioritized using a queue and searching is cut off
heuristically~\cite{DBLP:conf/sisap/BoytsovN13,kgraph,10.1007/978-3-319-46759-7_2,DBLP:journals/corr/abs-1810-07355,faiss,Muja09fastapproximate,Hajebi:2011:FAN:2283516.2283615,DBLP:journals/corr/abs-1802-02422,DBLP:journals/corr/MalkovY16,MALKOV201461}. %
KNNG based methods tend to perform better than tree-based and bucketing search
procedures on dense learned embeddings, as we study in this paper.
Unfortunately, KNNG search methods usually do not provide a formal \emph{proof}
for the quality of the return results.  These methods still include a tunable
parameter $\epsilon$ to control stopping heuristics which trade-off recall and
runtime.  KNNG methods, like bucketing methods, are unable to be provable
\emph{exact} without having to search the entire nearest neighbor graph.

\subsection{Certified Cosine (\cc)}

In this paper we introduce \emph{Certified Cosine} (\cc), a novel approach for
generating \emph{certificates} for fast nearest neighbor methods.  \cc builds on
prior KNNG based ANN techniques to search for nearest
neighbors. However, unlike prior probabilistic and heuristic approaches, \cc
constructs a \defn{certificate} which \underline{guarantees} that the nearest
neighbor returned is 100\% correct ($\epsilon = 0$).
This allows for \cc to be \emph{exact} and \emph{fast} when a certificate is
successfully constructed.  Unfortunately, certificates can not always be efficiently
constructed.  In this case, depending on the needs of an application, a user of
\cc can choose to either use the current best guess $\hat{v}$---which is akin
to current ANN methods---or request that the result is exact and
perform a linear scan over all vectors.

In \cref{sec:define_nn}, we begin by defining equivalent definitions of what it
means to be the nearest neighbor, which can then be used to construct a
certificate.  In \cref{sec:cert_strategies}, we discuss exactly how we implement
our certification strategy such that it is tractable to process while
simultaneously performing a fast nearest neighbor search.  Finally, in
\cref{sec:experiment_results}, we demonstrate that the additional overhead
introduced by our certification processes is manageable and that \cc achieves
query runtime performance that is comparable to the current state-of-the-art
approximate nearest neighbor methods.

\section{Definition of 1-Nearest Neighbor} \label{sec:define_nn}

Here we introduce the major definitions that we will use throughout this paper.
Given that \cc builds on the KNNG, we adopt
similar terminology to \citet{Sebastian:2002:MSR:839291.842892}
and NGT~\cite{10.1007/978-3-319-46759-7_2,DBLP:journals/corr/abs-1810-07355} as
both of these use an \emph{exact} KNNG in their search procedure as we do
here.

We will explain \cc as searching for and certifying the 1-\emph{nearest
  neighbor} for ease of presentation.  However note, \cc can be easily
generalized to the certify top-$k$ nearest neighbor set.\footnotemark{}%

We start by defining the \defn{query} $q \in \vecspace$ as the target vector
and our dataset $\vertices \subset \vecspace$ as a set of discrete vectors
$\vi \in \vertices$ in our vector space.  For convenience, we additionally
define $v^*$ as the \defn{\emph{true} 1-nearest neighbor}
($v^* := \text{argmin}_{\vi \in \vertices} d(\vi, q)$).  For reasons that will
become apparent, our method's \defn{distance function} is specific to cosine
similarity $d(x,y) = 1 - \text{cosine}(x, y)$.%
\footnotetext{\label{footnote:how_to_do_topk} To prove the top-$k$ nearest
  neighbors note, our definition (to follow) of $\mathcal{N}_{\hat{v}}$ only
  requires that we know the distance between our current best guess $\hat{v}$
  and the query $q$.  As such, to prove the top-$k$ nearest neighbors, we
  instead check the entire region of $\mathcal{N}_{v^{(k)}}$ contains exactly
  $k-1$ vectors rather than being an empty set,
  $|\mathcal{N}_{v^{(k)}} \cap \mathcal{V}| = k-1$.  This, in turn, means that
  there can not be a better vector located within this region that we have 
  observed (and included in the top-$k$.}%
\footnote{It is possible to preprocess the data such that Euclidean
  ($\text{argmin } \norm{\vi - q}$) or maximal inner product
  ($\text{argmax } \vi^T q$) can be converted to cosine and used as the distance
  metric as shown in \citet{Bachrach:2014:SUX:2645710.2645741}.}
Given that our distance metric is cosine similarity, we will assume that
all vectors are unit norm, which allows us to write cosine similarity as an inner
product between two vectors ($\text{cosine}(x,y) = x^T y$).
\begin{equation}
  v^* := \underset{\vi \in \mathcal{V}}{\text{argmin }} 1 - \text{cosine}(\vi, q) %
  \equiv \underset{\vi \in \mathcal{V}}{\text{argmax }} \vi^T q
  \label{eqn:vstar_def}
\end{equation}

Our search index is based on an exact \defn{$\mathbf{K}$ nearest neighbor graph}
(\defn{KNNG})~\cite{Arya:1993:ANN:313559.313768,Sebastian:2002:MSR:839291.842892}
$G = (\vertices, \lneighbor)$.  KNNG is a directed graph with vertices being the
vectors from the original dataset $\vertices$ and edges $\lneighbor$ being the
top-$k$ nearest neighbor to a vertex according to our distance metric, cosine
similarity.  The KNNG is constructed once during a preprocessing phase and then
reused for every query requiring $O(nk)$ to store, as is typical with ANN
methods.  We denote the edge set of each vertex as $\lneighbor_i$ which contains
the $k$ nearest neighbors.  The choice of $k$ impacts the \emph{success} rate of
constructing certificates vs. storage and search efficiency.

\begin{figure}[t]
\begin{figure}[H]
  \centering
    \centering
    \begin{tikzpicture}
    \def\angle{55}
    \def\pangle{45}
    \def\angleqp{105}
    \def\sangoff{7}

    \def\lcolor{purple}

    \clip (-2.2,-.7) rectangle(2.2,2.37) ;

    \draw (0,0) circle (2cm) ;
    \node (q) at (0.05,2.2) {$q$} ;
    \draw[thick,->] (0,0) to (0,2) ;
    \node[circle,fill,inner sep=1pt] (x1) at ({cos(\angle)*2}, {sin(\angle)*2}) {} ;
    \node at ([xshift=.5cm,yshift=.2cm]x1) {$v^* \in \vertices$} ;
    \draw[thick,->] (0,0) to (x1.center) ;

    \draw[\lcolor,dashed,-] (({-cos(\angle)*2}, {sin(\angle)*2}) to (({cos(\angle)*2}, {sin(\angle)*2}) ;

    \draw[thick,\lcolor] (({cos(-\angle)*2.05}, {sin(\angle)*2.05}) arc ({\angle}:{180-\angle}:2.05cm) ;
    \node at ({cos(180-(\angle+\pangle/4-\sangoff))*2.3},{sin(180-(\angle+\pangle/4-\sangoff))*2.3}) {\color{\lcolor}$\mathcal{N}_{v^*}$} ;

    \node[circle,fill,inner sep=1pt] (x2) at ({cos(150)*2}, {sin(150)*2}) {};
    \node at ([xshift=.4cm]x2) {$\v{r}$} ;

    \node[circle,fill,inner sep=1pt] (x3) at ({cos(25)*2}, {sin(25)*2}) {};
    \node at ([xshift=.2cm,yshift=.2cm]x3) {$\v{s}$} ;

    \node[circle,fill,inner sep=1pt] (x4) at ({cos(345)*2}, {sin(345)*2}) {};
    \node at ([xshift=-.2cm,yshift=.2cm]x4) {$\v{u}$} ;

  \end{tikzpicture}

  \caption{The query's neighborhood $\mathcal{N}_{v^*}$ of $q$ defined as
    $\mathcal{N}_{v^*} := \{ x \in \vecspace : q^T x > q^T v^* \}$ when
    intersected with the dataset $\mathcal{V}$ contains no points as there is
    nothing better than $v^*$.}
  \label{fig:intro_nn}
\end{figure}
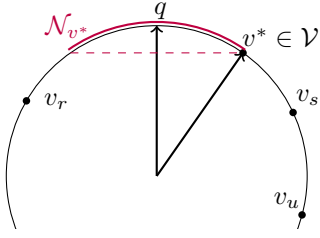
\begin{figure}[H]
  \centering
  \begin{tikzpicture}
    \def\points{
      a/165,
      b/155,
      c/143,
      d/138,
      e/122,
      f/105,
      g/95,
      h/80,
      i/65,
      j/55,
      k/40,
      l/32,
      m/23,
      n/16,
      o/-4,
      q/-13
    }

    \clip (-3.5,-.7) rectangle(3.5,2.37) ;
    
    \draw(0,0) circle (2cm) ;

    \foreach \xname/\xang in \points
    {
      \node[circle,fill,inner sep=1pt]  (x\xname) at ({cos(\xang)*2}, {sin(\xang)*2}) {};
      \node[]  (n\xname) at ({cos(\xang)*2.4}, {sin(\xang)*2.3}) {$\v{\xname}$};
    }

    \def\angle{55}  %
    \def\pangle{50}  %
    \draw[thick,->] (0,0) to (xj.center) ;
    \draw[thick,blue,-,fill=gray,fill opacity=.2] (({cos(\angle-\pangle)*2}, {sin(\angle-\pangle)*2}) -- (({cos(\angle+\pangle)*2}, {sin(\angle+\pangle)*2}) --
    (({cos(\angle+\pangle)*2}, {sin(\angle+\pangle)*2}) arc ({\angle+\pangle}:{\angle-\pangle}:2cm) -- cycle ;

    \def\lcolor{purple}
    \draw[thick,\lcolor] (({cos(\angle+\pangle)*2.1}, {sin(\angle+\pangle)*2.1}) arc ({\angle+\pangle}:{\angle-\pangle}:2.1cm) ;

    \node[] (ball_point) at ({cos(\angle-\pangle+5)*2.1}, {sin(\angle-\pangle+5)*2.1}) {};

    \node[\lcolor,xshift=.6cm] (ball_label) at ({cos(\angle-\pangle)*2.3}, {sin(\angle-\pangle)*2.3}) {\color{\lcolor}$\overline{\mathcal{B}}_{b_j}(v_j)$} ;

    \draw[->,shorten <= -.16cm,\lcolor,thick] (ball_label) -- (ball_point.center) ;

    \def\angle{143}  %
    \def\pangle{63}  %
    \draw[thick,->] (0,0) to (xc.center) ;
    \draw[thick,blue,-,fill=gray,fill opacity=.2] (({cos(\angle-\pangle)*2}, {sin(\angle-\pangle)*2}) -- (({cos(\angle+\pangle)*2}, {sin(\angle+\pangle)*2}) --
    (({cos(\angle+\pangle)*2}, {sin(\angle+\pangle)*2}) arc ({\angle+\pangle}:{\angle-\pangle}:2cm) -- cycle ;

  \end{tikzpicture}
  \caption{Neighborhoods of $\v{c}$ and $\v{j}$ constructed during preprocessing
    with all points with cosine $\ge b_c, b_j$ are contained.
    $b_c = \v{h}^T \v{c}$, $b_j = \v{f}^T \v{j}$,
    $\lneighbor_c = \{ a, b, d, e, f, g, h \}$,
    $\lneighbor_j = \{ f, g, h, i, k, l, m, n \}$.  }

  \label{fig:vertex_neighborhood}

\end{figure}
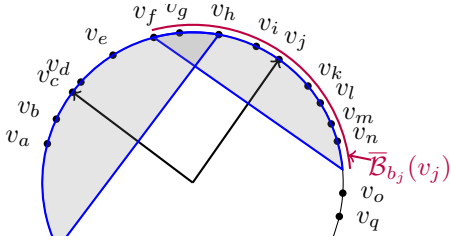
\end{figure}

For certification, we additionally define the \defn{neighborhood} around a point
and its associated size $b_i$.  A neighborhood is constructed such that we know
for certain that all vertices within a distance $b_i$ are contained in the
neighborhood:%
\begin{equation}
  \forall \v{j} \in \vertices, \;\; \vi^T \v{j} \ge b_i \implies \v{j} \in \lneighbor_i.
  \label{eqn:point_neighborhood}
\end{equation}
This, in turn, lets us define a ball of size $b_i$ centered at $\vi$ that is
entirely contained inside of the neighborhood and thus can serve as a compact
\emph{summary} of $\vi$'s edge set:
$\overline{\mathcal{B}}_{b_i}(\vi)\cap\vertices~\subseteq~\lneighbor_i$.%
\footnote{The ball $\mathcal{B}$ here is defined as usual, but we use the cosine distance:
  $\overline{\mathcal{B}}_r(v) := \{ x \in \vecspace : v^T x \ge r \}$ and
  $\mathcal{B}_r(v) := \{ x \in \vecspace : v^T x > r \}$.} %
Observe, given that $\lneighbor_i$ is constructed via an exact KNNG, $b_i$
simply becomes the distance of the $k^{\text{th}}$ nearest neighbor as show in
\cref{fig:vertex_neighborhood}.

We further define $\hat{v}$ as our current \defn{best guess} and
$\mathcal{N}_{\vi}$ as the \defn{query's neighborhood} parameterized by some
$\vi$ as $\mathcal{N}_{\vi} := \mathcal{B}_{\vi^T q}(q)$ (\Cref{fig:intro_nn}).
Observe, when $\mathcal{N}_{\hat{v}}$ is parameterize by the current best guess, it
represents the space that \emph{might} contain better $\vi$.  Only in the case
that $\mathcal{N}_{\vi}$ is parameterized by the \emph{true} 1-nearest neighbor,
will its intersection with the vertices be empty:
\begin{equation}
  \mathcal{N}_{\hat{v}} \cap \vertices = \emptyset \iff \hat{v} = v^*.
  \label{eqn:intersect_equiv}
\end{equation}

\subsection{What are Certificates?} \label{sec:what_are_cert}

\begin{figure}
  \centering
  \begin{subfigure}[t]{.48\textwidth}
    \centering

  \begin{tikzpicture}
    \def\angle{55}
    \def\pangle{55}
    \def\angleqp{105}
    \def\sangoff{7}

    \clip (-2.2,-.2) rectangle(2.2,2.37) ;

    \draw (0,0) circle (2cm) ;
    \node (q) at (0.1,2.2) {$q$} ;
    \draw[thick,->] (0,0) to (0,2) ;
    \node (x1) at ({cos(\angle)*2}, {sin(\angle)*2}) {} ;
    \node at ([xshift=.2cm,yshift=.2cm]x1) {$\hat{v}$} ;
    \draw[thick,->] (0,0) to (x1.center) ;

    \draw[red,dashed,-] (({-cos(\angle)*2}, {sin(\angle)*2}) to (({cos(\angle)*2}, {sin(\angle)*2}) ;

    \draw[thick,blue,-,fill=gray,fill opacity=.2] (({cos(\angle-\pangle)*2}, {sin(\angle-\pangle)*2}) -- (({cos(\angle+\pangle)*2}, {sin(\angle+\pangle)*2}) --
    (({cos(\angle+\pangle)*2}, {sin(\angle+\pangle)*2}) arc ({\angle+\pangle}:{\angle-\pangle}:2cm) -- cycle ;

    \draw[thick,green] (({cos(\angle+\pangle)*2.05}, {sin(\angle+\pangle)*2.05}) arc ({\angle+\pangle}:{180-\angle}:2.05cm) ;
    \node at ({cos(180-(\angle+\pangle/4-\sangoff))*2.3},{sin(180-(\angle+\pangle/4-\sangoff))*2.3}) {\color{green}$S_1$} ;

  \end{tikzpicture}

  \caption{First, $\hat{v}$ located that is near the query point $q$.  It is
    insufficient to prove that we have found the nearest neighbor as $S_1$ is
    non-empty indicating that there is some space where a better neighbor may
    lie.}

\end{subfigure}%

\begin{subfigure}[t]{.48\textwidth}
    \centering
    \begin{tikzpicture}
    \def\angle{55}
    \def\pangle{55}
    \def\angleN{140}

    \clip (-2.2,-.2) rectangle(2.2,2.37) ;

    \draw (0,0) circle (2cm) ;
    \node (q) at (0.1,2.2) {$q$} ;
    \draw[thick,->] (0,0) to (0,2) ;
    \node (x1) at ({cos(\angle)*2}, {sin(\angle)*2}) {} ;
    \node at ([xshift=.2cm,yshift=.2cm]x1) {$\hat{v}$} ;
    \draw[thick,->] (0,0) to (x1.center) ;

    \draw[red,dashed,-] (({-cos(\angle)*2}, {sin(\angle)*2}) to (({cos(\angle)*2}, {sin(\angle)*2}) ;

    \draw[thick,blue,-,fill=gray,fill opacity=.2] (({cos(\angle-\pangle)*2}, {sin(\angle-\pangle)*2}) -- (({cos(\angle+\pangle)*2}, {sin(\angle+\pangle)*2}) --
    (({cos(\angle+\pangle)*2}, {sin(\angle+\pangle)*2}) arc ({\angle+\pangle}:{\angle-\pangle}:2cm) -- cycle ;

    \node (x2) at ({cos(\angleN)*2}, {sin(\angleN)*2}) {} ;
    \node at ([xshift=-.2cm,yshift=.2cm]x2) {$v_{j}$} ;
    \draw[thick,->] (0,0) to (x2.center) ;

    \draw[thick,blue,-,fill=gray,fill opacity=.2] (({cos(\angleN-\pangle)*2}, {sin(\angleN-\pangle)*2}) -- (({cos(\angleN+\pangle)*2}, {sin(\angleN+\pangle)*2}) --
    (({cos(\angleN+\pangle)*2}, {sin(\angleN+\pangle)*2}) arc ({\angleN+\pangle}:{\angleN-\pangle}:2cm) -- cycle ;

  \end{tikzpicture}
  \caption{$v_{j}$ completes the proof that $\hat{v}$ (the first point that we
    located) is the nearest neighbor.  $S_2 = \emptyset$ which indicates that we
    have checked everywhere a better 1-nearest neighbor might lie.}

\end{subfigure}
\caption{Successful proof that $\hat{v}$ is the 1-nearest neighbor to $q$
  requiring two steps before we have checked \emph{all} of the space where a
  better neighbor might lie.}

  \label{fig:2d_proof}
\end{figure}
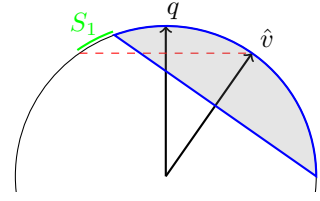
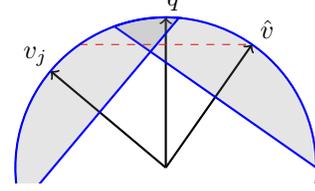

A certificate needs to \underline{guarantee} that $\hat{v} = v^*$, which means
that we need some way to check this statement or an equivalent statement.
\Cref{eqn:intersect_equiv} introduced an equivalence between checking for an
empty set intersection and constructing a certificate for $\hat{v}$.
The question remains \emph{how} to efficiently check this intersection is
empty. A simple strategy would be to perform a linear scan over all of
$\mathcal{V}$.  Given that $\mathcal{V}$ is a discrete set, this is tractable.
However, a linear scan over all of the data is exactly what we are trying to
avoid as a \emph{fast} NN method.

Rather, we are going to start with the \emph{assumption} that $\hat{v} = v^*$
and will search for counterexamples to this assumption.  A certificate is then
the case where we have proven that a counterexample can not exist.
To identify where a counterexamples $v' \in \mathcal{V}$ \emph{might} exist, we
define the \defn{unchecked region} (\Cref{fig:2d_proof}) starting with
$S_0 := \mathcal{N}_{\hat{v}}$ with successive unchecked regions shrinking,
$S_t \subseteq S_{t-1}$.  We will momentarily define how we shrink $S_t$.  Now,
if $\hat{v}$ is \emph{not} the true 1-nearest neighbor, then there must exist a
$v' \in S_t \cap \vertices$ as a counterexample.  As such, once we prove that
$S_t = \emptyset$, then it is impossible for there to exist $v' \in S_t$ in which
case $\hat{v}$ \underline{must} be the true 1-nearest neighbor.

To construct successive smaller unchecked regions, we only have to check the
difference $S_{t-1} - S_t$ which we are removing at every step from $S_{t-1}$.  This follows
from a short proof on the sequence $S_t$:
\begin{align}
  S_t = \emptyset \implies& S_t \cap \mathcal{V} = S_{t-1} \cap \mathcal{V} =
  \ldots = S_0 \cap \mathcal{V} = \emptyset \nonumber \\
  \equiv& \; \mathcal{N}_{\hat{v}} \cap \mathcal{V} = \emptyset \iff \hat{v} = v^*
  \label{eqn:St_sequence}
\end{align}
Where this chain of equalities follows from \cref{eqn:St_chain} where we restrict
the region that we are checking to $\vertices$ and thus only have discrete
points that need to be checked.
\begin{equation}
  \mathcal{V} \cap (S_{t-1} - S_t) = \emptyset \iff
  \lnot \exists v' \in \mathcal{V} \text{ s.t. } v' \in S_{t-1} - S_t.
  \label{eqn:St_chain}
\end{equation}
When checking $\vertices \cap (S_{t-1} - S_t)$, if we find $v'$ then that implies a contradiction
with the original assumption $\hat{v} = v^*$.
\begin{align}
  \exists v' \in \mathcal{V} \text{ s.t. } v' \in S_{t-1} - S_t \implies v' \in S_0 \cap \mathcal{V} \nonumber  \\
  \equiv v' \in  \mathcal{N}_{\hat{v}} \cap \mathcal{V} \implies \hat{v} \neq v^* \qquad \square
\end{align}

\emph{Procedurally}, when finding $v' \in \mathcal{V} \cap S_t$ \cc restarts the
certification process using $v'$ as the new guess for the 1-nearest neighbor
($\hat{v} \gets v'$).

To check only this subset $\vertices \cap (S_{t-1} - S_t)$ without having to
scan all of $\vertices$, we make use of the KNNG and the fact that we have
preprocessed the neighborhood around every vertex in the graph.  Essentially,
for some $\v{j}$ that is selected during \cc's search procedure, we will have
$S_{t-1} - S_{t} \subseteq \overline{\mathcal{B}}_{b_j}(\v{j})$.  Thus, we can
check the neighborhood of $\v{j}$ to mark this area as checked
$(S_{t-1} - S_t) \cap \mathcal{V} \subseteq \lneighbor_j$.  Checking
$\lneighbor_j$ is easy since it is a small set of size $k$ for which we can check
all vectors referenced.

All that we need now to complete \cc's certification process is an efficient way
to track $S_t$ and identify when this set is empty.

\section{Tracking the Unchecked Region $S_t$} \label{sec:cert_strategies}

Our eventual goal is to prove $S_t = \emptyset$ as that indicates that a
certificate has been successfully constructed.  To make this tractable, we
essentially want a compact \emph{summary} of where we have searched.  Now, when
searching for the nearest neighbor, once we have checked all of the vectors
referenced in a neighbor set $\lneighbor_i$, then we know that we have fully
searched the neighborhood around $\v{i}$ (\cref{eqn:point_neighborhood}).  This
in turn means that we can summarize the searched area as:
$\{ x : x^T \v{i} \ge b_i \}$.  This follows from the fact that we are using
cosine similarity as our distance metric which allows us to represent the
distance using an inner product (\cref{eqn:vstar_def}), and that we know that
all vectors within distance $b_i$ of $\v{i}$ \emph{must} be contained within
$\lneighbor_i$.

To make this more concrete, we define a \defn{constraint store} $\constraints$,
which tracks the regions that is still unchecked.  Any time that we have
completed processing the neighborhood $\lneighbor_j$, we add the constraint
$\constraints_t \gets \constraints_{t-1} \cup \{ \{ x : x^T \vi \le b_i \} \}$,
which represents the area that we have not checked.  We can now track $S_t$ as
the intersection of $\constraints$ and the subspace $\vecspace$ as follows:
\begin{align}
  S_t = \{ x : \norm{x} = 1 \} \cap \{ x : x^T q \ge q^T \hat{v} \} \; \cap \nonumber \\
  \left( \bigcap_{\{x : x^T \vi \le b_i\} \in \constraints_t} \{x : x^T \vi \le b_i\} \right)
\end{align}

We can easily handle most of these constraints, as $q$ and $\vi$ are constants,
which makes these linear constraints.  However, the surface of the sphere,
$\norm{x} = 1$, is non-convex and thus requires special handling.  To actually
implement checking if $S_t$ is empty, we employ a number of different strategies
to check and relaxations of the above non-convex relaxations that we will
cover in the next sections.

\subsection{Single Point Certificate}
\label{sec:single_point_cert}

\begin{figure}[H]
  \centering
  \begin{tikzpicture}
    \clip (-1.6,1.2) rectangle (1.6,2.4) ;
    \def\angle{65}
    \def\pangle{55}
    \def\angleqp{105}
    \def\sangoff{7}

    \draw (0,0) circle (2cm) ;
    \node (q) at (0.1,2.2) {$q$} ;
    \draw[thick,->] (0,0) to (0,2) ;
    \node (x1) at ({cos(\angle)*2}, {sin(\angle)*2}) {} ;
    \node at ([xshift=.2cm,yshift=.2cm]x1) {$\hat{v}$} ;
    \draw[thick,->] (0,0) to (x1.center) ;

    \draw[red,dashed,-] (({-cos(\angle)*2}, {sin(\angle)*2}) to (({cos(\angle)*2}, {sin(\angle)*2}) ;

    \draw[thick,blue,-,fill=gray,fill opacity=.2] (({cos(\angle-\pangle)*2}, {sin(\angle-\pangle)*2}) -- (({cos(\angle+\pangle)*2}, {sin(\angle+\pangle)*2}) --
    (({cos(\angle+\pangle)*2}, {sin(\angle+\pangle)*2}) arc ({\angle+\pangle}:{\angle-\pangle}:2cm) -- cycle ;

  \end{tikzpicture}
  \caption{Single point certificate, $\mathcal{N}_{\hat{v}}$ is entirely
    contained inside $\overline{\mathcal{B}}_{\hat{b}}(\hat{v})$.}

  \label{fig:single_proof}
\end{figure}
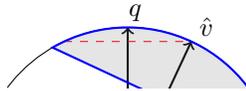

First, the easiest case is where we can prove that $S_1 = \emptyset$ with a
single neighbor.  This occurs when the distance between the
query $q$ and $\vi$ is sufficiently close.  Then, it is possible that the query
neighborhood will be completely contained inside of neighborhood of $\vi$ as
shown in \cref{fig:single_proof}
($\mathcal{N}_{\hat{v}} \subseteq \overline{\mathcal{B}}_{b_i}(\vi)$).  The
check for this case is simply $\acos(q^T \hat{v}) + \acos(q^T \vi) < \acos(b_i)$
where we check if the angle between $\vi$ and $q$ plus the angle defining the
query neighborhood for certification ($\acos(q^T \hat{v})$) fits inside of the
angle that corresponds with $\vi$'s neighborhood size ($\acos(b_i)$).

\subsection{Convex Relaxation} \label{sec:project}

Our main goal is to determine if $S_t$ is empty.  If we are unable to do that
with the single point certificate, then we to have to use multiple points to
determine if $S_t$ is empty.  Our general approach is to identify a
counterexample to the claim that $S_t = \emptyset$ by finding a $x \in S_t$.  If
we have a \emph{complete} method that only fails to find $x \in S_t$ if and only
if $S_t = \emptyset$ then we can use failure to locate $x$ as the certificate.

Unfortunately, it is in general difficult to directly locate $x \in S_t$ as it
is non-convex and potentially not even connected due to constraint
$\norm{x} = 1$.
Instead, we use a convex relaxation of $S_t$ by including the inside of the unit
ball $\norm{x} \le 1$.
With all convex constraints, this problem is known as the \emph{convex
  feasibility problem} where we are trying to determine if a set defined as the
intersection of multiple convex sets is empty.  This problem can be reduced to
locating a point inside of the intersection of all of the
sets~\cite{Bauschke:1996:PAS:240441.240442}.  To solve this, we use alternating
projection (\Cref{algo:alt_project}) as it has low overhead, is easy to
implement, and provides guarantees of finding some point inside of the
intersection if it is non-empty~\cite{Bauschke1993}. We can additionally alter
the order and frequency we check constraints for better efficiency.
In the case that the intersection of constraints is empty (also implying
$S_t = \emptyset$), then the sequence generated by alternating projection does
not converge but rather oscillates between different points in the sets that we
are intersecting.  When we detect this case, we report that the intersection is
empty and the certificate is complete (\Cref{line:alt_check_converge} in
\cref{algo:alt_project}).

\begin{align}
  Proj&_{c} (x) = x - \vi *\max(0, \vi^T x - b_i) \label{eqn:proj_vi} \\
  Proj&_{\hat{v}} (x) = x + q *\max(0, q^T\hat{v} - q^T x) \label{eqn:proj_pd} \\
  Proj&_{\norm{x}\le 1} (x) = \frac{x}{\max(\norm{x}, 1)} \label{eqn:proj_norm}
\end{align}

\begin{algorithm}[H]
  \begin{algorithmic}[1]
    \Function{SolveProject}{$x$, \constraints}
      \State \Comment{Returns $\langle$$S_t$ is empty, $x \in \text{conv}(S_t)\rangle$ \hfill}
      \For{$t \in 1,2,3,\ldots$}
        \State $x_o \gets x$
        \For{$c \in \constraints$}
          $x \gets Proj_c(x)$ \Comment{Eqn. (\ref{eqn:proj_vi})}
        \EndFor
        \State $x \gets Proj_{x^T q \ge \hat{v}^T q}(x)$ \Comment{Eqn. (\ref{eqn:proj_pd})}
        \State $x \gets Proj_{\norm{x} \le 1}(x)$ \Comment{Eqn. (\ref{eqn:proj_norm})}
        \If{$x = x_o$}
          \Return $\langle \text{false}, x \rangle$
        \EndIf
        \If{$\norm{x - x_o} > \alpha^t$}
          \Return $\langle \text{true}, \_ \rangle$ \label{line:alt_check_converge}
        \EndIf
      \EndFor
    \EndFunction
  \end{algorithmic}
  \caption{Alternating projection loops over all constraints until we identify a
    point in the intersection.}
  \label{algo:alt_project}
\end{algorithm}

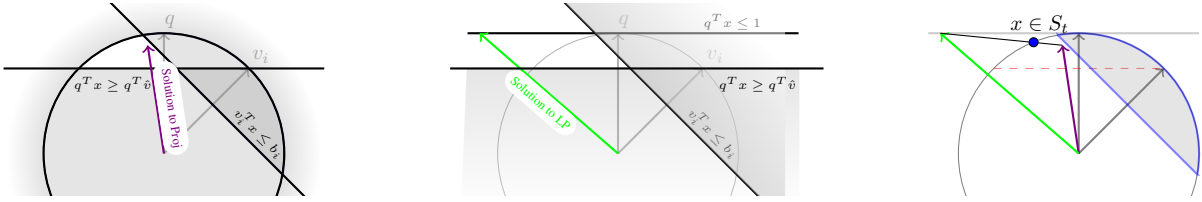
\begin{figure*}
  \begin{subfigure}[t]{.30\textwidth}
    \begin{tikzpicture}[scale=0.8, every node/.style={scale=0.8}]
      \def\angle{45}
      \def\pangle{55}
      \def\angleqp{105}
      \def\sangoff{7}
      
      \clip (-3.5,-.7) rectangle(3.5,2.5) ;
    
      \filldraw[inner color=gray,outer color=white!100,white]
      (0,0) circle (2.6cm) ;
      \draw[fill=white] (0,0) circle (2cm) ;

      \begin{scope}[opacity=.3,name prefix=org-]

        \draw (0,0) circle (2cm) ;
        \node (q) at (0.1,2.2) {$q$} ;
        \draw[thick,->] (0,0) to (0,2) ;
        \node (x1) at ({cos(\angle)*2}, {sin(\angle)*2}) {} ;
        \node at ([xshift=.2cm,yshift=.2cm]x1) {$\v{i}$} ;
        \draw[thick,->] (0,0) to (x1.center) ;

        \node (redA) at (({-cos(\angle)*2}, {sin(\angle)*2}) {};
        \node (redB) at (({cos(\angle)*2}, {sin(\angle)*2}) {} ;
        \draw[red,dashed,-] (redA.center) to (redB.center) ;

        \node (areaA) at (({cos(\angle-\pangle)*2}, {sin(\angle-\pangle)*2}) {} ;
        \node (areaB) at (({cos(\angle+\pangle)*2}, {sin(\angle+\pangle)*2}) {} ;

        \draw[thick,blue,-,fill=gray,fill opacity=.2] (areaA.center) -- (areaB.center) --
        (({cos(\angle+\pangle)*2}, {sin(\angle+\pangle)*2}) arc ({\angle+\pangle}:{\angle-\pangle}:2cm) -- cycle ;

      \end{scope}

      \draw[fill=gray,fill opacity=.2] (org-redA.center) -- (org-redB.center) --
      (org-redB.center) arc (\angle:{-360+\angle*3}:2cm)  ;

      \draw[thick,-,shorten >=-1cm,shorten <=-1cm] (org-areaA) -- node[pos=.2,above=-.05cm,sloped] {\tiny $\vi^T x \le b_i$} (org-areaB) ;

      \draw[thick] (0,0) circle (2cm) ;
      \draw[thick,-,shorten >=-1cm,shorten <=-1cm] (org-redA.center) -- node[pos=.2,below] {\tiny $q^Tx \ge q^T\hat{v}$} (org-redB.center) ;

      \draw[thick,violet,->] (0,0) -- node[pos=.4,above,sloped,fill=white,rounded corners=5pt] {\tiny Solution to Proj} (-.27,1.8) ;

    \end{tikzpicture}
    \caption{Convex relaxation of the projection operator. \Cref{sec:project}}
    \label{fig:project}
  \end{subfigure}%
  \qquad
\begin{subfigure}[t]{.3\textwidth}
  \begin{tikzpicture}[scale=0.8, every node/.style={scale=0.8}]
    \clip (-3.5,-.7) rectangle(3.5,2.5) ;

    \draw[white] (0,0) circle (2.6cm) ; %

    \begin{scope}[opacity=.3,name prefix=org-]

        \def\angle{45}
        \def\pangle{55}
        \def\angleqp{105}
        \def\sangoff{7}

        \draw (0,0) circle (2cm) ;
        \node (q) at (0.1,2.2) {$q$} ;
        \draw[thick,->] (0,0) to (0,2) ;
        \node (x1) at ({cos(\angle)*2}, {sin(\angle)*2}) {} ;
        \node at ([xshift=.2cm,yshift=.2cm]x1) {$\v{i}$} ;
        \draw[thick,->] (0,0) to (x1.center) ;

        \draw[red,dashed,-] (({-cos(\angle)*2}, {sin(\angle)*2}) to (({cos(\angle)*2}, {sin(\angle)*2}) ;

        \node (redA) at (({-cos(\angle)*2}, {sin(\angle)*2}) {};
        \node (redB) at (({cos(\angle)*2}, {sin(\angle)*2}) {} ;

        \node (areaA) at (({cos(\angle-\pangle)*2}, {sin(\angle-\pangle)*2}) {} ;
        \node (areaB) at (({cos(\angle+\pangle)*2}, {sin(\angle+\pangle)*2}) {} ;

      \end{scope}
      \draw[white,shading=axis,opacity=.2] ([xshift=-1.1cm]org-redA.center) -- ([xshift=1.6cm]org-redB.center) --
      ([xshift=1.6cm,yshift=-2.4cm]org-redB.center) --
      ([xshift=-1.2cm,yshift=-2.4cm]org-redA.center) -- cycle ;

      \draw[thick,-] (-2.5,2) -- node[pos=.8,above=-.05cm] {\tiny $q^Tx \le 1$} (3,2) ;
      \draw[thick,-,shorten >=-1cm,shorten <=-1.3cm] (org-areaA) -- node[pos=-.45] (ll1) {} node[pos=1.35] (ll2) {}  node[pos=.2,above=-.05cm,sloped] {\tiny $\vi^T x \le b_i$} (org-areaB) ;

      \draw[shading=axis,shading angle=135,fill opacity=.5,draw opacity=0]
      let \p1 = (ll1.center),
          \p2 = (ll2.center)
      in
      (\p1) -- (\p2) -- (\x1,\y2) -- cycle;

      \draw[thick,->,green] (0,0) -- node[pos=.5,below,sloped,fill=white,rounded corners=5pt] {\tiny Solution to LP} (-2.3,2) ;

      \draw[thick,-,shorten >=-1.7cm,shorten <=-1.2cm] (org-redA) -- node[pos=1.4,below] {\tiny $q^T x \ge q^T \hat{v}$} (org-redB) ;

  \end{tikzpicture}
  \caption{Convex relaxation to a linear program. \Cref{sec:lp}}
  \label{fig:lp}
\end{subfigure}
\qquad
\begin{subfigure}[t]{.3\textwidth}
  \begin{tikzpicture}[scale=0.8, every node/.style={scale=0.8}]
    \def\angle{45}
    \def\pangle{55}
    \def\angleqp{105}
    \def\sangoff{7}
    
    \clip (-3.5,-.7) rectangle(3.5,2.5) ;

    \draw[white] (0,0) circle (2.6cm) ; %

      \begin{scope}[opacity=.5,name prefix=org-]

        \draw (0,0) circle (2cm) ;
        \draw[thick,->] (0,0) to (0,2) ;
        \node (x1) at ({cos(\angle)*2}, {sin(\angle)*2}) {} ;
        \draw[thick,->] (0,0) to (x1.center) ;

        \node (redA) at (({-cos(\angle)*2}, {sin(\angle)*2}) {};
        \node (redB) at (({cos(\angle)*2}, {sin(\angle)*2}) {} ;
        \draw[red,dashed,-] (redA.center) to (redB.center) ;

        \node (areaA) at (({cos(\angle-\pangle)*2}, {sin(\angle-\pangle)*2}) {} ;
        \node (areaB) at (({cos(\angle+\pangle)*2}, {sin(\angle+\pangle)*2}) {} ;

        \draw[thick,blue,-,fill=gray,fill opacity=.2] (areaA.center) -- (areaB.center) --
        (({cos(\angle+\pangle)*2}, {sin(\angle+\pangle)*2}) arc ({\angle+\pangle}:{\angle-\pangle}:2cm) -- cycle ;

      \end{scope}

      \draw[thick,violet,->] (0,0) -- (-.27,1.8) ;

      \draw[thick,-,opacity=.2] (-2.5,2) -- (2,2) ;

      \draw[thick,->,green] (0,0) --  (-2.3,2) ;

      \draw[-] (-2.3,2) -- (-.27,1.8) ;

      \node[circle,draw,fill=blue,inner sep=1.5pt] (x) at (-.75,1.85) {};
      \node at ([xshift=.1cm,yshift=.3cm]x) {$x\in S_t$} ;

  \end{tikzpicture}
  \caption{Combining the results of projection and linear program to find
    $x \in S_t$. \Cref{sec:combine}}
  \label{fig:combine}
\end{subfigure}
\caption{Converting $S_t$ into two convex sets which we can find points inside.
  We can then find $x \in S_t$ by drawing a line through $S_t$ and finding the
  point that has $\norm{x} = 1$.  The shaded regions represent what has been
  eliminated by our constraints \constraints.}
w\label{fig:conv_st}
\end{figure*}

\subsection{Linear Programming Relaxation}
\label{sec:lp}

We can choose to relax $S_t$ by removing the norm constraint, leaving only
linear constraints.  By choosing to use the original query $q$ and maximize the
objective $q^T x$ with the constraint $q^T x \le 1$, we can be certain that we
will find $x^T q \ge \hat{v}^T q$ if it exists.  As such, if
$x^T q < \hat{v}^T q$ then $S_t = \emptyset$ as
$S_t \subset \{ x : x^T q \ge \hat{v}^T q \}$ and by solving the linear
programming relaxation, we can find a solution such that we are confident there
is nothing better.  Off the shelf simplex~\cite{simplex_method} solvers are
optimized for large problems with many sparse constraints, whereas here we have
a small number of dense constraints.  In support of our experiments, we
implement a custom solver optimized for this condition.
By not enforcing $\norm{x} \le 1$ and using simplex, the solver usually locates a
sparse solution that is far outside of the unit ball, \cref{fig:lp}.

\begin{align}
  \max_{x \in \mathbb{R}^n} \; q^T x \\
  q^Tx &\le 1 \nonumber \\
  \forall \{ x : \vi^T x \le b_i \} \in \constraints, \;\; \vi^T x &\le b_i \nonumber 
\end{align}

\subsection{Finding a Counterexample, $x \in S_t$}
\label{sec:combine}

If we are merely interested in checking if $S_t$ is empty, then using either the
convex or linear programming relaxation is sufficient.  However, when tracking
if $S_t$ is empty, we are also finding counterexamples in the convex
relaxations.  We can use these points (both outside and inside of the unit ball)
to locate a point on the surface of the unit ball inside of the original
non-convex $S_t$.  We can further use these new points to retarget our search
towards areas which we have not explored to avoid getting ourselves stuck in a dense,
well-connected cluster (\Cref{sec:finding_guesses}).
To do this, we can draw a line between the result from the
projection method and the linear program to find the point
along the line that has unit norm (\Cref{fig:combine,eqn:line_between}).
\begin{align}
  x_{\in S_t} = x_{proj} + \frac{x_{lp} - x_{proj}}{\norm{x_{lp} - x_{proj}}} * \nonumber \qquad\qquad\qquad \\
  \sqrt{(1 - \norm{x_{proj}}^2)(1 - \cosine(x_{proj}, x_{lp})^2)} \label{eqn:line_between}
\end{align}

\section{Finding Good Guesses $\hat{v}$} \label{sec:finding_guesses}

\begin{algorithm*}[t]
  \begin{algorithmic}[1]
    \Function{Lookup}{$q$, budget}
    \State count $\gets 0$; certified $\gets$ false; $x \gets q$
    \State $E \gets \{\}$ \Comment{All processed vertices}
    \State $Q \gets \{ $ \Call{Seed}{$x$} $ \}$ \Comment{Priority queue of unexpanded vertices ordered by $x^Tv$}
    \State $\constraints \gets \{ \}$ \Comment{Set of constraints used for certification, \Cref{sec:cert_strategies}}
      \While{$|Q|$ $>$ 0 \textbf{ and } count++ $<$ budget \textbf{ and not } certified} \label{algo:line:stopping}
        \State $v_j \gets \text{argmax}_{v' \in Q} \; x^T v'$ \Comment{Select the closest unfinished vertex from the priority queue} \label{algo:line:select_queue}
        \State $v_n \gets \text{argmax}_{v' \in \lneighbor_j - E} \; v_j^T v'$ \Comment{Select next unprocessed neighbor from $\lneighbor_j$}
        \State $E \gets E \cup \{ \v{n} \}$; $Q \gets Q \cup \{ \v{n} \}$ \Comment{Record expanded item and add to priority queue}
        \If{$\hat{v}^T q < \v{n}^T q$}
          $\hat{v} \gets \v{n}$ \Comment{Update $\hat{v}$ with new 1 best located}
        \EndIf
        \If{$\lneighbor_j \subset E$} \Comment{Completed processing the entire neighbor set of $\v{j}$}
          \State $Q \gets Q - \{ v_j \}$; $\constraints \gets \constraints \cup \{ \{ x : x^T v_j \le b_j \} \}$ \Comment{Remove $\v{j}$ from queue and add constraint for tracking $S_t$}
          \State $\langle x, \text{certified} \rangle \gets $ \Call{ConstructCertificate}{$q$, \constraints, $\v{j}$, $b_j$} \Comment{Try constructing certificate that $S_t = \emptyset$ or find $x \in S_t$} \label{algo:line:new_x}
        \EndIf
      \EndWhile
      \State \Return $\langle \hat{v}, \text{certified} \rangle$ \Comment{Return the best $\vi$ found during searching}
    \EndFunction

    \Function{ConstructCertificate}{$q$, \constraints, $\v{j}$, $b_j$}
    \If{$\acos(\v{j}^Tq) + \acos(\hat{v}^T q) < \acos(b_j)$}
      \Return $\langle \_, \text{true} \rangle$ \Comment{\Cref{sec:single_point_cert}}
    \EndIf

    \State $\langle x_{proj}, \text{emptyIntersection} \rangle \gets$ \Call{SolveProject}{$q$, \constraints} \Comment{\Cref{sec:project,algo:alt_project}}
    \If{emptyIntersection}
      \Return $\langle \_, \text{true} \rangle$
    \EndIf

    \If{$x_{proj} / \norm{x_{proj}} \in S_t$}
      \Return $\langle x_{proj} / \norm{x_{proj}}, \text{false} \rangle$
    \EndIf

    \State $x_{lp} \gets $ \Call{SolveSimplex}{$q$, \constraints} \Comment{\Cref{sec:lp}}
    \If{$x_{lp}^T q < \hat{v}^T q$}
      \Return $\langle \_, \text{true} \rangle$
    \EndIf
    \If{$\norm{x_{lp}} < 1$}
      \Return $\langle \_, \text{false} \rangle$ \Comment{Can not prove $S_t = \emptyset$ and can not find $x \in S_t$}
    \EndIf

    \State $x = x_{proj} + \frac{x_{lp} - x_{proj}}{\norm{x_{lp} - x_{proj}}} \sqrt{(1 - \norm{x_{proj}}^2)(1 - \cosine(x_{proj}, x_{lp})^2)}$ \Comment{\Cref{sec:combine}}
    \State \Return $\langle x, \text{false} \rangle$

    \EndFunction

  \end{algorithmic}
  \caption{Outline of our lookup procedure intermixed with constructing
    certificates.  The point $x \in S_t$ is continuously adjusted to
    towards under-explored regions to better target search (\Cref{sec:finding_guesses}).}
  \label{algo:lookup_prove}
\end{algorithm*}

\cc certification procedure requires that we first find a good guess $\hat{v}$
before we can attempt certification.  We employ techniques similar to other ANN
KNNG based searched
methods~\cite{DBLP:journals/corr/abs-1810-07355,10.1007/978-3-319-46759-7_2,kgraph}.
In practice, \cc's ANN search procedure could be based on any ANN method, though
using a KNNG for search allows us to reuse operations between searching and
certification.  \cc's search procedure, \cref{algo:lookup_prove}, uses a
priority queue to track the current nearest unexplored neighbors.  We initialize
the priority queue with a single seed vector that is selected using LSH using
the first $m$ sign bits~\cite{Indyk:1998:ANN:276698.276876,Charikar:2002}.  \cc
eagerly jumps to the current best vector as it is located
(\cref{algo:lookup_prove} \cref{algo:line:select_queue}).  When a vertex's
neighbor set $\lneighbor_i$ is fully explored, \cc add $\v{i}$'s constraint to
$\constraints$, in turn shrinking $S_t$.  Tracking $S_t$ also help the
search procedure better target its efforts.  By locating a $x \in S_t$, (e.g.
$\text{argmax}_{x \in S_t} x^T q$), \cc re-prioritizes its search towards areas
that are currently unexplored.  This helps quickly find vertices which can help
with constructing a certificate (\cref{algo:lookup_prove}
\cref{algo:line:new_x}).  \cc only terminates its search when a certificate is
successfully constructed or when a prespecified budget has been exceeded.

\section{Experimental  Results} \label{sec:experiment_results}

\begin{figure*}
  {\centering
  \begin{subfigure}[t]{.49\textwidth}
    \centering
    \begin{tikzpicture}[scale=0.86, ]  
  \begin{axis}[
    name=ax1,
    xlabel={ Recall },
    ylabel={ Queries per second (1/s) },
    ymode = log,
    yticklabel style={/pgf/number format/fixed,
      /pgf/number format/precision=3},
    legend style = { anchor=west, at={(6.5cm,3cm)}},
    mark size=1pt
    ]
    \addplot[color=Blue,mark=*,mark size=.8pt] coordinates {
      (0.12453, 824.324448353)
      (0.22821, 223.154501318)
      (0.33168, 108.407387995)
      (0.38225, 47.0484597091)
      (0.49974, 44.9388357614)
      (0.58713, 24.6464898274)
      (0.68075, 18.7760359059)
      (0.78436, 18.7726366024)
      (0.89763, 15.13377773)
      (0.93767, 13.0313358508)
      (0.95793, 11.7892331105)
      (0.98312, 8.83109791286)
    };
    \addlegendentry{ BallTree(nmslib) };

    \addplot[color=Red,mark=*,mark size=1.8pt,dashed] coordinates {
      (1.0, 10.720359984)
    };
    \addlegendentry{ bruteforce-blas };

    \addplot[color=Emerald,mark=*,mark size=.8pt,densely dotted] coordinates {
      (0.10401, 26235.0437)
      (0.16877, 17369.5812512)
      (0.25822, 15136.623933)
      (0.35518, 12526.7907249)
      (0.37075, 9797.06090789)
      (0.42911, 9539.71157315)
      (0.47777, 8140.99001937)
      (0.49609, 6950.89928406)
      (0.55308, 5931.10797624)
      (0.59639, 5146.67576329)
      (0.61589, 4540.65764613)
      (0.66224, 3720.35925031)
      (0.66696, 3385.93312145)
      (0.6987, 3095.51503157)
      (0.71696, 2816.44596593)
      (0.74905, 2263.11309922)
      (0.7509, 2232.29039599)
      (0.76092, 2094.41993392)
      (0.78321, 1699.32271057)
      (0.79677, 1665.82494458)
      (0.79934, 1491.92475112)
      (0.80207, 1488.24186088)
      (0.83273, 1224.75344571)
      (0.83471, 1203.23895104)
      (0.83744, 1064.65338398)
      (0.85108, 986.116699511)
      (0.86652, 878.61400705)
      (0.87539, 776.552279648)
      (0.88317, 703.869876009)
      (0.89697, 612.930848452)
      (0.90175, 566.244341646)
      (0.91232, 473.078067902)
      (0.91702, 452.067484364)
      (0.91851, 423.652927546)
      (0.92942, 392.476157913)
      (0.94034, 305.904619406)
      (0.94258, 301.766999032)
      (0.95383, 243.132753997)
      (0.9581, 211.442514072)
      (0.9638, 210.090125788)
      (0.96913, 167.931945231)
      (0.97336, 161.011860013)
      (0.97782, 145.009231983)
      (0.97838, 128.940673922)
      (0.98512, 111.456910447)
      (0.99061, 85.6629803483)
    };
    \addlegendentry{ hnswlib };
    \addplot[color=Purple,mark=square*,mark size=.8pt] coordinates {
      (1.0, 1.47839708681)
    };
    \addlegendentry{ kd };
    \addplot[color=Salmon,mark=square*,mark size=.8pt] coordinates {
      (0.86889, 409.624009407)
      (0.87496, 387.457739891)
      (0.88196, 369.041528716)
      (0.88749, 349.853139895)
      (0.89342, 334.389203261)
      (0.91275, 267.060517359)
      (0.93643, 197.453835153)
      (0.94972, 158.057341874)
      (0.95851, 132.79247844)
      (0.9653, 114.807594488)
      (0.97035, 101.87688582)
      (0.97415, 91.977778525)
      (0.97759, 84.1110568905)
      (0.97966, 77.4903377901)
      (0.98215, 72.2833950462)
    };
    \addlegendentry{ kgraph };
    \addplot[color=LimeGreen,mark=square*,mark size=.8pt,dashed] coordinates {
      (0.0119, 39073.4833805)
      (0.07813, 25440.2685288)
      (0.18656, 16384.4864144)
      (0.49326, 7113.90445331)
      (0.58538, 5289.78785503)
      (0.73887, 2938.84767124)
      (0.93882, 549.625977409)
      (0.99957, 31.3977149476)
    };
    \addlegendentry{ NGT-onng };
    \addplot[color=CadetBlue,mark=triangle*,mark size=1.1pt] coordinates {
      (0.01177, 40515.2032085)
      (0.07244, 25479.6001803)
      (0.16103, 16228.9564113)
      (0.42263, 6593.90939235)
      (0.51463, 4774.38184546)
      (0.65785, 2810.33829357)
      (0.83943, 1104.20449754)
      (0.98198, 108.888061375)
    };
    \addlegendentry{ NGT-panng };
    \addplot[color=Maroon,mark=triangle*,mark size=1.1pt,dashed] coordinates {
      (0.00347, 23589.5468874)
      (0.00458, 15110.3944341)
      (0.01764, 4863.37616752)
      (0.22408, 329.896940948)
      (0.53545, 58.357504306)
      (0.80313, 18.9620647116)
    };
    \addlegendentry{ rpforest };
    \addplot[color=Rhodamine,mark=triangle*,mark size=1.1pt,densely dotted] coordinates {
      (0.05278, 2035.33372468)
      (0.06543, 1904.38068435)
      (0.08311, 1699.09450984)
      (0.08952, 1623.92191151)
      (0.11321, 1348.97993761)
      (0.14801, 1064.70981399)
      (0.17103, 901.783908461)
      (0.1893, 789.666845556)
      (0.22029, 642.850126494)
      (0.58939, 613.653239739)
      (0.64126, 508.818264733)
      (0.69941, 388.82086253)
      (0.7655, 371.4754657)
      (0.81302, 275.199675569)
      (0.85031, 201.16628468)
      (0.87798, 149.925199754)
      (0.8916, 126.799463163)
      (0.90811, 101.714130394)
      (0.92366, 79.4354816999)
      (0.93439, 65.7229821924)
      (0.94428, 53.6968076932)
      (0.9522, 46.3463083488)
      (0.95815, 40.6647558435)
      (0.96042, 38.2700272186)
    };
    \addlegendentry{ SW-graph(nmslib) };

    \addplot[mark=diamond*,mark options={fill=red},mark size=1.6pt,thick] coordinates {
      (0.43139, 3830.2561547)
      (0.43889, 3565.70361391)
      (0.66642, 1294.09149051)
      (0.68212, 1257.47135383)
      (0.68852, 1232.39176005)
      (0.76768, 786.884128866)
      (0.77874, 772.928884087)
      (0.79574, 668.765061271)
      (0.80569, 657.915683753)
      (0.86926, 402.413480776)
      (0.87026, 396.399190796)
      (0.88822, 341.668375767)
      (0.92736, 204.514991184)
      (0.93696, 173.122502634)
      (0.94855, 136.693555801)
      (0.95669, 115.53805922)
      (0.97075, 82.0577050094)
      (0.97636, 69.2655567141)
      (0.99032, 33.3560143623)
      (0.99266, 26.905735612)
    };
    \addlegendentry{ \bf This Paper };

    \legend{} ;
  \end{axis}



\end{tikzpicture}
    \caption{GloVe~\cite{pennington2014glove} 200 dim 1,183,514 entries}
    \label{fig:plot:recall_compare}
  \end{subfigure}%
  \hfill
  \begin{subfigure}[t]{.52\textwidth}
    \hspace{5mm}
\begin{tikzpicture}[scale=0.86]
  \begin{axis}[
    xlabel={ Recall },
    ymode = log,
    yticklabel style={/pgf/number format/fixed,
      /pgf/number format/precision=3},
    legend style = { anchor=west, at={(6.5cm,3cm)}},
    mark size=1pt
    ]
    \addplot[color=Red,mark=*,mark size=1.8pt,dashed] coordinates {
      (0.9991, 36.3526971491)
    };
    \addlegendentry{ bruteforce-blas };

    \addplot[color=Emerald,mark=*,mark size=.8pt,densely dotted] coordinates {
      (0.10224, 29246.0246349)
      (0.17003, 19125.4546184)
      (0.29249, 16948.4701215)
      (0.40408, 13608.8741339)
      (0.41595, 11467.0311209)
      (0.47535, 10607.8671589)
      (0.53171, 8824.51801089)
      (0.55024, 8138.43416156)
      (0.59597, 6822.5857743)
      (0.59747, 6350.51163033)
      (0.64143, 5472.15794577)
      (0.65249, 5244.44481122)
      (0.68929, 4243.1387077)
      (0.72222, 3321.40306993)
      (0.72936, 3240.56491453)
      (0.75531, 2539.55783048)
      (0.75896, 2480.52450361)
      (0.76028, 2208.28818842)
      (0.77523, 1944.19751022)
      (0.78753, 1895.9570223)
      (0.78915, 1655.94120406)
      (0.79343, 1597.31913567)
      (0.81066, 1424.63564714)
      (0.81459, 1354.37497606)
      (0.81538, 1295.00363356)
      (0.81709, 1183.39501693)
      (0.82464, 1100.76075584)
      (0.83685, 1000.50092982)
      (0.83792, 925.237064887)
      (0.8443, 860.169239259)
      (0.85072, 771.71934907)
      (0.86006, 719.065223986)
      (0.8637, 645.910361686)
      (0.86478, 630.399489801)
      (0.88348, 499.442341715)
      (0.8896, 404.207387923)
      (0.89564, 327.337801574)
      (0.9071, 310.25159714)
      (0.90744, 251.878089776)
      (0.9101, 222.543876394)
      (0.91616, 210.195269179)
      (0.91947, 172.227665597)
      (0.92817, 162.9117682)
      (0.93549, 112.940902966)
      (0.93896, 87.5950380352)
    };
    \addlegendentry{ hnswlib };

    \addplot[color=Salmon,mark=square*,mark size=.8pt] coordinates {
      (0.9167, 1288.67313832)
      (0.92265, 1208.09201687)
      (0.92583, 1150.7838454)
      (0.93008, 1072.46891836)
      (0.932, 1037.23315531)
      (0.9409, 856.622962796)
      (0.95516, 637.156672792)
      (0.96297, 511.240166806)
      (0.96935, 425.752756049)
      (0.97396, 364.036994657)
      (0.97719, 319.854048532)
      (0.98056, 284.001535637)
      (0.98261, 255.735030787)
      (0.98464, 233.358233645)
      (0.9861, 214.5953065)
    };
    \addlegendentry{ kgraph };
    \addplot[color=LimeGreen,mark=square*,mark size=.8pt,dashed] coordinates {
      (0.05342, 48306.0033768)
      (0.10723, 39164.9158585)
      (0.19856, 24573.1443319)
      (0.52861, 10309.2299018)
      (0.65582, 7514.5110565)
      (0.85979, 4529.78484865)
      (0.97552, 1030.41554428)
      (0.99848, 87.5492849633)
    };
    \addlegendentry{ NGT-onng };
    \addplot[color=CadetBlue,mark=triangle*,mark size=1.1pt] coordinates {
      (0.05332, 46539.7217805)
      (0.10649, 37349.4664685)
      (0.18598, 21716.8869955)
      (0.50278, 8550.03610161)
      (0.62905, 6296.10787931)
      (0.78435, 4096.74413517)
      (0.91214, 1947.82221025)
      (0.99005, 140.465825936)
    };
    \addlegendentry{ NGT-panng };

    \addplot[color=Purple,mark=square*,mark size=.8pt] coordinates {
      (0.40628, 9.08795899796)
    };
    \addlegendentry{ kd };

    \addplot[mark=diamond*,mark options={fill=red},mark size=1.6pt,thick] coordinates {
      (0.16245, 19220.415239)
      (0.1653, 18243.2192033)
      (0.25588, 13012.2189344)
      (0.27675, 12433.8488036)
      (0.58604, 3611.62531853)
      (0.75701, 3225.38062923)
      (0.82949, 2280.48906801)
      (0.83253, 1297.53280913)
      (0.90213, 1236.82766377)
      (0.9069, 1227.84384059)
      (0.90921, 1192.54908168)
      (0.93072, 791.545562431)
      (0.93438, 791.535703486)
      (0.93745, 759.196421501)
      (0.94052, 698.183580841)
      (0.94343, 663.665357154)
      (0.95558, 491.092043977)
      (0.95815, 461.540963677)
      (0.95976, 432.721166674)
      (0.96262, 404.784882224)
      (0.9707, 309.762138236)
      (0.97367, 286.411194471)
      (0.97833, 230.029481949)
      (0.98093, 210.750217501)
      (0.98606, 161.820683933)
      (0.98658, 156.579868574)
      (0.98795, 148.568925176)
      (0.98852, 143.669240464)
      (0.9921, 109.607783491)
      (0.99392, 101.258711725)
      (0.99519, 88.4875863341)
    };
    \addlegendentry{ \bf This Paper };

    \legend{} ;

  \end{axis}
\end{tikzpicture}
    \hspace{2mm} \;
    \caption{NYTimes~\cite{Dua:2019} 256 dim 290,000 entries}
    \label{fig:plot:recall_compare2}
  \end{subfigure}
  \
  }

  \begin{minipage}[b]{.6\textwidth}
    \caption{Recall-Queries per second (1/s) tradeoff of top 10 - up and to the
      right is better.  To sweep out \cc's time vs. recall we adjust the budget
      of how many vertices we can expand ranging from 1000 to 50000 and
      controlling $K$, the number of neighbors in the KNNG graph.  Plots are
      generated by ANN-Benchmark~\cite{ann-benchmark}.  Experiments were run on
      an Intel E5-2667 v3.}
  \label{fig:plot:ann_recall}
\end{minipage}%
\qquad
  \begin{minipage}[b]{.2\textwidth}
    \scalebox{.65}{
\begin{tikzpicture}

  \begin{axis}[%
    width=1.7cm,  
    height=1.6cm,
    hide axis,
    xmin=0,
    xmax=1,
    ymin=0,
    ymax=0.4,
    legend style={draw=white!15!black,at={(0cm,0cm)}},
    ]

    \addlegendimage{color=Blue,mark=*,mark size=1.8pt}
    \addlegendentry{ BallTree(nmslib)~\cite{Yianilos:1993:DSA:313559.313789,DBLP:conf/sisap/BoytsovN13} };

    \addlegendimage{color=Red,mark=*,mark size=1.8pt,dashed}
    \addlegendentry{ bruteforce-BLAS (linear scan) };

    \addlegendimage{color=Emerald,mark=*,mark size=1.8pt,densely dotted}
    \addlegendentry{ hnswlib~\cite{DBLP:journals/corr/abs-1802-02422} };

    \addlegendimage{color=Purple,mark=square*,mark size=1.8pt}
    \addlegendentry{ kd-tree~\cite{kdtrees} };

    \addlegendimage{color=Salmon,mark=square*,mark size=1.8pt}
    \addlegendentry{ kgraph~\cite{kgraph} };

    \addlegendimage{color=LimeGreen,mark=square*,mark size=1.8pt,dashed}
    \addlegendentry{ NGT-onng~\cite{DBLP:journals/corr/abs-1810-07355} };

    \addlegendimage{color=CadetBlue,mark=triangle*,mark size=2.1pt}
    \addlegendentry{ NGT-panng~\cite{10.1007/978-3-319-46759-7_2} };

    \addlegendimage{color=Maroon,mark=triangle*,mark size=2.1pt,dashed}
    \addlegendentry{ rpforest~\cite{rpforest} };

    \addlegendimage{color=Rhodamine,mark=triangle*,mark size=2.1pt,densely dotted}
    \addlegendentry{ SW-graph(nmslib)~\cite{DBLP:conf/sisap/BoytsovN13} };

    \addlegendimage{mark=diamond*,mark options={fill=red},mark size=2.6pt,thick}
    \addlegendentry{ \bf \cc (This Paper) };

  \end{axis}
\end{tikzpicture}
}
  \end{minipage}
  \vspace{20mm}
\end{figure*}
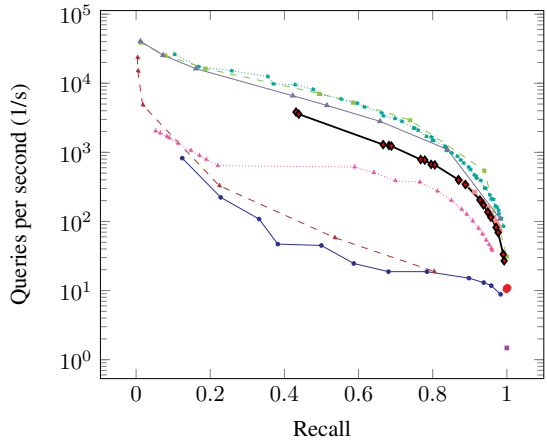
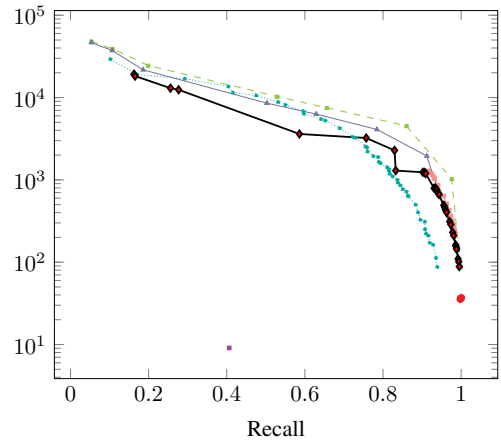

\begin{figure*}
  \begin{subfigure}[b]{.55\textwidth}
    \centering
    {
      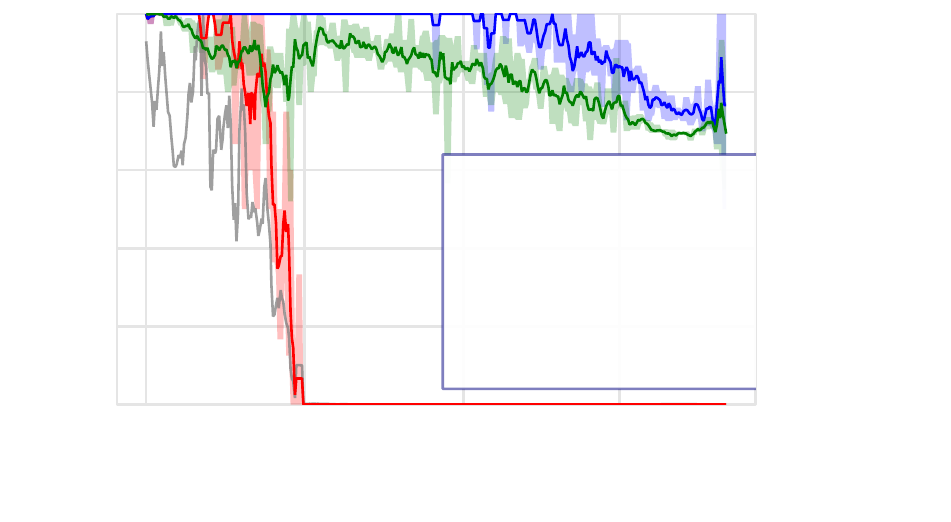
    }
    \vspace{-5mm}  %
    \caption{Rolling average recall and successful construction of a certificate
      over .001 sized buckets grouped by $d(v^*, q)$.  Certification was of
      \emph{only} of the top-1 nearest neighbor with a stopping criteria of
      either a successful certificate or a hard budget of expanding 50,000
      vertices.  Dataset is GloVe 200 dimension with 1,183,514 entries.}
    \label{fig:plot:dist_vs_recall}
  \end{subfigure}%
  \hfill
  \begin{subfigure}[b]{.38\textwidth}
    \vspace{0pt}
    {
\begin{tikzpicture}[scale=0.7, every node/.style={scale=1}]
  \begin{axis}[
    xlabel={ Recall },
    ylabel={ Queries per second (1/s) },
    ymode = log,
    yticklabel style={/pgf/number format/fixed,
      /pgf/number format/precision=3},
    legend style = { anchor=west, at={(6.5cm,3cm)}},
    mark size=1pt
    ]
    \addplot[color=Red,mark=*,mark size=1.8pt,dashed] coordinates {
      (1.0, 7.46948966593)
    };
    \addlegendentry{ bruteforce-blas };
    \addplot[color=Emerald,mark=*,mark size=.8pt,densely dotted] coordinates {
      (0.00943, 26702.3181076)
      (0.01784, 17314.3379283)
      (0.029595, 16566.748711)
      (0.05596, 11832.4979388)
      (0.07018, 9489.91339894)
      (0.087915, 7233.8085846)
      (0.091785, 7026.21329712)
      (0.106965, 5759.27297264)
      (0.11494, 4997.23855265)
      (0.131385, 4367.93532592)
      (0.137055, 4221.74343253)
      (0.156075, 3417.50509248)
      (0.167465, 3010.2396351)
      (0.18665, 2561.8327956)
      (0.19704, 2501.31676374)
      (0.220285, 1995.39954349)
      (0.2276, 1838.17565918)
      (0.2363, 1785.74050758)
      (0.261765, 1455.19677933)
      (0.27242, 1449.83046267)
      (0.28637, 1191.22919538)
      (0.32385, 1045.81842519)
      (0.331625, 937.268709563)
      (0.35496, 843.511068083)
      (0.381875, 744.464681053)
      (0.397945, 680.125538767)
      (0.41942, 605.160490593)
      (0.43542, 514.089719485)
      (0.46396, 483.247440905)
      (0.509985, 391.171390537)
      (0.514, 376.545357421)
      (0.552505, 275.473807371)
      (0.59365, 261.788833544)
      (0.650045, 210.127216395)
      (0.67627, 179.739555734)
      (0.736145, 143.666101323)
      (0.795315, 109.220914365)
      (0.82695, 59.9832236481)
    };
    \addlegendentry{ hnswlib };
    \addplot[color=Salmon,mark=square*,mark size=.8pt] coordinates {
      (0.27081, 657.47857575)
      (0.27701, 624.634925375)
      (0.29265, 582.108983569)
      (0.30029, 559.18556912)
      (0.311615, 529.147659029)
      (0.35388, 422.012549194)
      (0.41978, 312.243800811)
      (0.4721, 247.009025258)
      (0.51143, 201.425838862)
      (0.547055, 174.586269428)
      (0.57889, 113.683042064)
      (0.60435, 102.970606614)
      (0.6297, 93.0104287292)
      (0.6517, 86.2405223236)
      (0.67197, 78.9696371345)
    };
    \addlegendentry{ kgraph };

    \addplot[color=Purple,mark=square*,mark size=.8pt] coordinates {
      (1.0, 1.7075711077)
    };
    \addlegendentry{ kd };
    
    \addplot[color=LimeGreen,mark=square*,mark size=.8pt,dashed] coordinates {
      (0.00141, 38374.9382422)
      (0.007865, 30094.0820338)
      (0.021535, 18981.5554198)
      (0.09076, 6645.21203526)
      (0.13474, 4216.14065731)
      (0.24586, 1758.96517606)
      (0.65829, 241.334989341)
      (0.999745, 10.0128924001)
    };
    \addlegendentry{ NGT-onng };
    \addplot[color=SkyBlue,mark=triangle*,mark size=1.1pt] coordinates {
      (0.001405, 39062.529977)
      (0.00677, 30914.2898207)
      (0.017865, 19554.812914)
      (0.072215, 6859.70523559)
      (0.10431, 4412.32236338)
      (0.18003, 2058.64701477)
      (0.401195, 488.804724127)
      (0.951, 22.592980217)
    };
    \addlegendentry{ NGT-panng };

    \addplot[mark=diamond*,mark options={fill=red},mark size=1.6pt,thick] coordinates {
      (0.002705, 22718.1295571)
      (0.00653, 22278.5348012)
      (0.00668, 21792.8488889)
      (0.0069, 15575.9721094)
      (0.01663, 15042.3703967)
      (0.016715, 14818.4483022)
      (0.01673, 14444.6480156)
      (0.01708, 13751.4940224)
      (0.06127, 4486.91598233)
      (0.09627, 4248.89142007)
      (0.099705, 4036.32098397)
      (0.10135, 3761.13805546)
      (0.163725, 1604.80988724)
      (0.20585, 1507.82427327)
      (0.219455, 1506.37269962)
      (0.22727, 1426.41844976)
      (0.23372, 970.495769735)
      (0.285615, 914.018094939)
      (0.300665, 911.50545099)
      (0.305245, 862.232172098)
      (0.33016, 772.479845612)
      (0.33436, 730.125622378)
      (0.355225, 487.204128868)
      (0.428865, 458.7756463)
      (0.42972, 434.268478872)
      (0.463285, 387.824167257)
      (0.50475, 244.016411236)
      (0.57098, 229.389071414)
      (0.572925, 229.138154685)
      (0.608135, 193.287853319)
      (0.60931, 192.695219837)
      (0.66377, 152.164840828)
      (0.69758, 128.025402306)
      (0.70359, 127.809574149)
      (0.779535, 90.538980337)
      (0.80448, 76.0542786301)
      (0.815395, 76.0032772072)
      (0.92068, 37.0304183107)
      (0.93064, 36.8633651257)
      (0.961415, 24.0952584208)
      (0.96933, 23.8639876238)
    };
    \addlegendentry{ \bf This paper };

    \legend{};

  \end{axis}
\end{tikzpicture}
    }
    \vspace{5mm}  %
    \caption{ Recall-Queries per second (1/s) tradeoff - up and to the right is
      better, GloVe 200 dim 1,183,514 entries.  Queries generated by selecting a
      random $\vi$ and computing $q = \vi + \mathcal{N}(0, \epsilon I)$ for
      $\epsilon \in [0,1]$ uniformly.  Legend in \cref{fig:plot:ann_recall}.}
    \label{fig:plot:recall_compare_s}
  \end{subfigure}
  \caption{Here we experiment with queries of different distances from their
    nearest neighbor.  The query distribution has a large impact on
    performance.  The area of $\mathcal{N}_{v^*}$ grows very quickly at a rate
    of $O(d(v^*,q)^{\frac{n}{2} + 2})$ where $n$ the dimension is generally
    between 100 and 1000 with this plot at $n=200$.}
  \label{fig:plot:dist_v_recall}

\end{figure*}

To compare \cc, we use ANN-benchmark (ANNB)~\cite{ann-benchmark} as a standard
testing framework.
ANNB includes wrappers for existing state-of-the-art methods
as well as standard test sets and hyperparameter configurations for running
experiments.  Except for KD-trees~\cite{kdtrees}, ball
trees~\cite{Yianilos:1993:DSA:313559.313789,DBLP:conf/sisap/BoytsovN13} and
brute force linear scan, the prior work that we compare against only provides
\emph{probabilistic} guarantees of returning the correct answer.
\Cref{fig:plot:ann_recall} plots the runtime (queries per second) vs the recall
of the top-10 nearest neighbors.  To limit the maximum runtime, \cc has a
tunable budget parameter (\cref{algo:lookup_prove} \cref{algo:line:stopping})
which allows us to limit how many vertices \cc expands before returning the
current best guess.

In \cref{fig:plot:recall_compare}, \cc dominates the other exact methods.  \cc
run 3 to 30 times faster than
Ball-trees~\cite{Yianilos:1993:DSA:313559.313789,DBLP:conf/sisap/BoytsovN13} for
a similar recall.  The linear scan using BLAS achieves 10.7 queries per second
and KD-trees%
\footnote{Note: KD-trees runtime performance on this dataset isn't
  uncharacteristic given this is a high dimensional and dense
  embeddings~\cite{when_nn_is_meaningful}. We primarily include KD-trees as it
  is the canonical baseline for exact NN methods.}~\cite{kdtrees} achieves 1.5
queries per second both with 100\% recall.  \cc similarly achieved 26.9 queries
per second at 99.2\%.
For these experiments, 99.2\% was the maximum recall that \cc achieved. We did
not require that \cc construct a certificate for all returned results rather
using the best guess in the case that the budget was exceeded.

For the ANN methods that we compare against,
NGT's~\cite{DBLP:journals/corr/abs-1810-07355} search procedure is the closest
to \cc's as it also uses an exact KNNG to as its primary search data structure.
NGT, however, uses a stopping heuristic which allows it to stop searching much
earlier than \cc and thus greatly benefits it when a lower recall is acceptable.
With high recall ($>$ 99\%) \cc is 14\% slower than NGT, likely due to NGT
heuristic allowing it to stop earlier.  For lower recall, $<$ 95\%, other ANN
are up to 3 times faster than \cc.

These results with \cc represent a significant improvement over previously
available options for applications which require the exact NN set.  The factor
of 3 difference from SotA approximate methods also indicates that the ability to
heuristically stop earlier is helpful for performance, though this does run
counter to \cc ability to generate certificates.

\subsection{The Impact of $d(v^*, q)$ on Performance}

We observe that the distance between the query $q$ and the 1-nearest neighbor
$v^*$ has a significant impact on certification success and runtime performance,
as shown in \cref{fig:plot:dist_v_recall}.  The impact of $d(v^*,q)$ is an
interesting metric to study as it correlates with the \emph{entropy} of a query.
A low entropy query would have a small distance from its nearest neighbor (hence
a distribution over $\vertices$ is peaked at its 1-nearest neighbor).  In
\cref{fig:plot:recall_compare_s}, we kept the same underlying data as
\cref{fig:plot:recall_compare} but change the distribution of queries.  Now,
with more lower entropy queries---as we might expect from a well trained
model---\cc outperforms all other ANN methods for high recall.  \cc
certification ability allows it to stop earlier than the heuristics in these
cases as it \emph{knows} for certain that it has located the correct nearest
neighbors.

\section{Conclusion}

We introduced \emph{Certified Cosine} (\cc), a novel approach for certifying the
correctness of the nearest neighbor set.  To our knowledge, this is the first
time that a (sometimes) exact method has been demonstrated to work well on high
dimensional dense learned embeddings.  While constructing a certificate is not
always feasible, we believe that this approach can help in situations which
require correctness and are currently utilizing linear scans.  Additionally, we
have demonstrated that it is possible to use powerful constraint solvers
(\cref{sec:project,sec:lp}) inside of a nearest neighbor lookup while still
being competitive.  Future work may consider adapting \cc's certificate
construction process in designing better heuristics for ANN methods.
\bibliographystyle{aaai}
\bibliography{bibfile}

\end{document}